\documentclass[letterpaper, 10 pt, conference]{ieeeconf}  

\IEEEoverridecommandlockouts                              
\usepackage[utf8]{inputenc}
\usepackage[T1]{fontenc}
\usepackage{cite}
\usepackage{algorithm}%
\usepackage{algorithmic}
\usepackage{graphicx}
\usepackage{textcomp}
\usepackage{multirow}
\usepackage{amsmath}%
\usepackage{amssymb}
\usepackage{tabularx}%
\usepackage{float}%
\usepackage{svg}%
\usepackage{url}%
\usepackage{hyperref}
\usepackage{breakurl}

\title{RISE-iEEG: Robust to Inter-Subject Electrodes Implantation Variability iEEG Classifier}
\author{Maryam Ostadsharif Memar, Navid Ziaei, Behzad Nazari, and Ali Yousefi%
\thanks{Maryam Ostadsharif Memar, Navid Ziaei, and Behzad Nazari are with the Department of Electrical and Computer Engineering, Isfahan University of Technology, Isfahan, Iran (e-mails: m.ostadsharif@ec.iut.ac.ir, n.ziaei@ec.iut.ac.ir, nazari@iut.ac.ir).}%
\thanks{Ali Yousefi is with the Department of Biomedical Engineering, University of Houston, Houston, Texas, USA (e-mail: aliyousefi@uh.edu).}}

\begin{document}

\maketitle
\thispagestyle{empty}
\pagestyle{empty}

\begin{abstract}
Intracranial electroencephalography (iEEG) is increasingly used for clinical and brain-computer interface applications due to its high spatial and temporal resolution. However, inter-subject variability in electrode implantation poses a challenge for developing generalized neural decoders. To address this, we introduce a novel decoder model that is robust to inter-subject electrode implantation variability. We call this model RISE-iEEG, which stands for Robust to Inter-Subject Electrode Implantation Variability iEEG Classifier. RISE-iEEG employs a deep neural network structure preceded by a participant-specific projection network. The projection network maps the neural data of individual participants onto a common low-dimensional space, compensating for the implantation variability. In other words, we developed an iEEG decoder model that can be applied across multiple participants' data without requiring the coordinates of electrode for each participant. The performance of RISE-iEEG across multiple datasets, including the Music Reconstruction dataset, and AJILE12 dataset, surpasses that of advanced iEEG decoder models such as HTNet and EEGNet. Our analysis shows that the performance of RISE-iEEG is about 7\% higher than that of HTNet and EEGNet in terms of F1 score, with an average F1 score of 0.83, which is the highest result among the evaluation methods defined. Furthermore, Our analysis of the projection network weights reveals that the Superior Temporal and Postcentral lobes are key encoding nodes for the Music Reconstruction and AJILE12 datasets, which aligns with the primary physiological principles governing these regions. This model improves decoding accuracy while maintaining interpretability and generalization.
\end{abstract}


\section{Introduction}
Researchers record brain activity using various techniques to understand brain functions and support neurological diagnoses and treatments \cite{ref46}. iEEG stands out for its balance of temporal and spatial resolution, initially used for epilepsy cases and now applied more broadly due to advances in neural interface technology. However, its invasive nature and ethical constraints limit access to consistent implantation coordinates across participants, leading to variable and non-overlapping brain coverage \cite{ref47}. Developing tools to handle this variability is essential for effectively utilizing iEEG data.

\begin{figure*}[t]
	\centering
	\includegraphics[width=\textwidth]{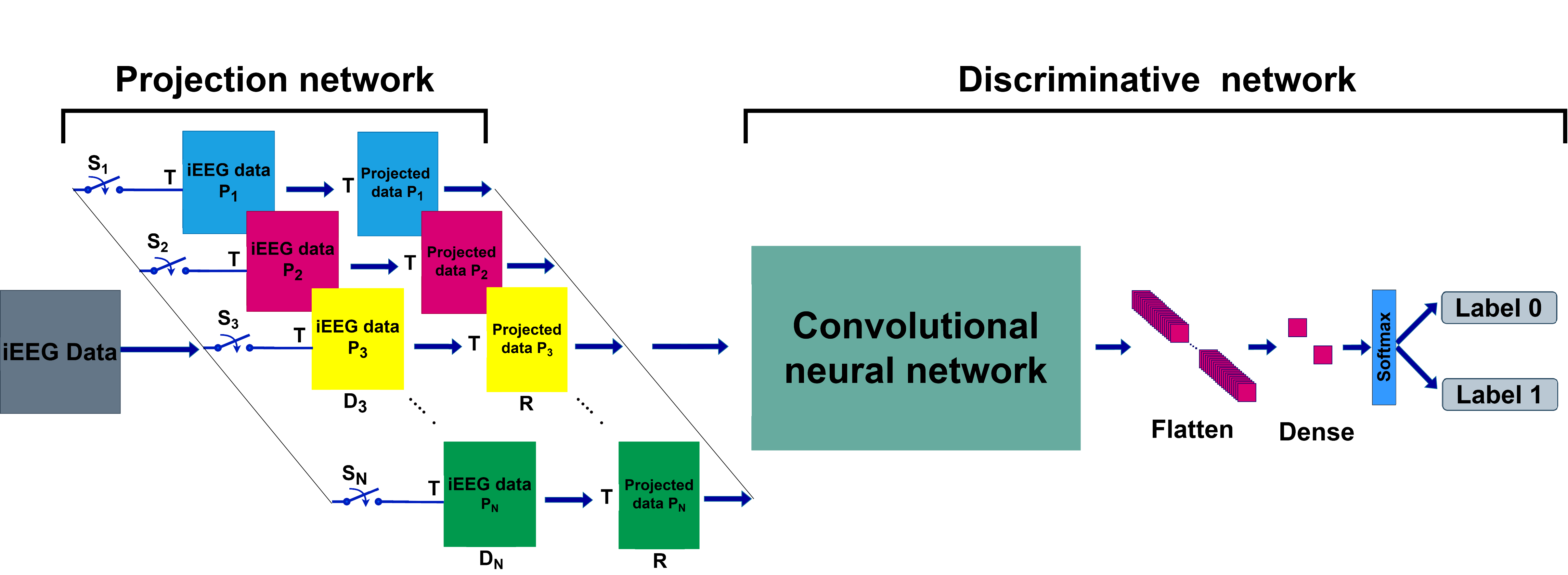}
	\caption{The RISE-iEEG model consists of two main networks: the projection network and the discriminative network. The projection network includes a participant-specific dense layer for each individual, which performs a linear transformation to map electrode data onto a common space. The discriminative network, based on the EEGNet architecture \cite{ref7}, extracts meaningful features from the projected data using temporal and spatial convolutional neural networks. In this figure, the notation $P_i$ represents participant $i$, and $S_i$ represents the switch control for participant $i$. The total number of participants is denoted by $N$, while $T$ indicates the number of sample times. The dimension of the common space is represented by $R$, and $D_i$ specifies the number of electrodes for participant $i$.}
	\centering
	\label{schematic model}
\end{figure*}

Beyond seizure prediction and localization \cite{ref32}, iEEG data is widely used in neural decoding and interface applications, such as decoding visual and audio-visual stimuli \cite{ref26}, \cite{ref27}, silent speech \cite{ref30}, and motor imagery \cite{ref31}. These tasks employ various methods, including traditional techniques like SVM, Bayesian linear discriminant analysis, and K-Nearest Neighbors \cite{ref66}, as well as newer approaches like the Bayesian Time-Series classifier \cite{ref26}. Recent deep learning models are increasingly used for decoding, offering end-to-end pipelines that bypass the need for preprocessing and feature extraction \cite{ref66}.

Deep learning models often require large datasets, making their application to iEEG challenging due to limited data and variations in electrode placement across participants. EEGNet \cite{ref7}, while effective for datasets like EEG-ImageNet \cite{ref37}, cannot handle inter-participant variability in electrode placement. Simple approaches like TCN \cite{ref60} and RNN \cite{ref61} concatenate electrode time series for generalization but struggle to identify meaningful patterns due to noisy or uninformative electrodes, reducing performance. The IEEG-HCT \cite{ref62} model processes data from multiple participants for basic tasks like artifact detection but is not suited for complex tasks like movement decoding or cognitive analysis, which require deeper understanding of brain dynamics.

A promising solution to this challenge is the HTNet model \cite{ref6}, which standardizes electrode positioning by mapping coordinates to predefined regions of interest (ROIs), enabling model training on data from various participants. However, this requires precise electrode positioning in MNI space, which involves the time-consuming and error-prone process of co-registering preoperative MRI with postoperative CT scans \cite{ref65}. Thus, many publicly available iEEG datasets lack MNI electrode data, hindering the mapping of electrode positions to a common space. Additionally, the mapping approach in this model relies on electrode distances from ROIs, which may not be suitable, as functional connections between brain areas aren’t based on physical distance \cite{ref41}. Given these challenges, developing new techniques to address electrode implantation variability is crucial for expanding iEEG's utility in clinical and neuroscience applications.

In this research, we introduce the RISE-iEEG model, designed to address the challenges of electrode implantation variability and to overcome the limitations of the approach used in the HTNet model. The RISE-iEEG model is built using projection and discriminative networks. The projection network maps the iEEG data of individual participants onto a common space using linear mappings, without requiring the MNI coordinates of electrodes. The weights of the projection network in the RISE-iEEG model are tuned during the training step. The discriminative network has an architecture similar to EEGNet, employing temporal and spatial convolutional neural networks to extract useful features from the data for classification. In this study, we discuss the RISE-iEEG architecture, its training process, and model assessment using two datasets. We also explore methods to interpret the trained model, aiming to investigate neural encoding mechanisms. We believe that post-training and performance analyses will validate the model architecture and demonstrate its utility in building a scalable and generalizable decoder model for iEEG and potentially EEG datasets. 

The paper is organized as follows: Section 2 explains the RISE-iEEG model, training, and cross-validation. Section 3 introduces the iEEG datasets, presents the decoding results, compares them with advanced decoders, and interprets the trained weights. Section 4 discusses advantages and limitations, and Section 5 provides the conclusion.

\section{Material and Methods}
\subsection{RISE-iEEG Model Architecture}
\label{model_archi}

The RISE-iEEG model is a cascade deep neural network comprising a participant-specific projection network followed by a discriminative deep neural network. The participant-specific projection network applies a linear transformation to map the input data onto a common lower-dimensional space, potentially addressing variability in electrode placement across participants. Then, the discriminative network, applied to the projected data from various participants, extracts key features to predict the corresponding labels.

The data for each participant \( i \) is represented as \( X^{(i)} \in \mathbb{R}^{M \times T \times D_i} \), where \( M \) is the number of trials, \( T \) is the number of time samples, and \( D_i \) is the number of electrodes for participant \( i \). While \( T \) and \( M \) are the same for all participants, \( D_i \) varies. The structure of the RISE-iEEG model is depicted in Fig. \ref{schematic model}.

The projection network ensures that subsequent layers are fed with a consistent representation of the iEEG data from various participants. This consistency enhances the network’s ability to extract robust features independent of individual variability. This mapping is expected to improve the model's generalization capabilities and accuracy in classifying neural patterns, regardless of variations in electrode placement across participants. As shown in Fig.\ref{schematic model}, the projection network includes participant-specific dense layers, which linearly map each participant's data onto a common space of dimension \( R \). The output of the projection layer for the \( j \)-th trial of participant \( i \), denoted as \( X_j^{(i)} \), is:

\begin{equation}
\begin{aligned}
    Z_i &= X_j^{(i)} W_{\text{Proj}}^{(i)} \in \mathbb{R}^{1 \times T \times R}, \\[1em]
    X_j^{(i)} &\in \mathbb{R}^{1 \times T \times D_i}, \quad W_{\text{Proj}}^{(i)} \in \mathbb{R}^{D_i \times R}
\end{aligned}
\tag{1}
\end{equation}
\vspace{0.5em}

Additionally, We have a participant selector \( s^i \in \{0,1\} \) in the projection network to identify which participant’s data is fed to the model in each iteration. Let \( S = [s^1, s^2, \dots, s^{N}] \) represent the participant selector, and \( Z = [z^1, z^2, \dots, z^{N}] \) denote the outputs of the dense layers. The output of the projection network is computed as the dot product of \( S \) and \( Z \):

\begin{equation}
Y_{\text{projection}} = S \cdot Z 
\tag{2}
\end{equation}
\vspace{0.5em}

The discriminative network is shared across all participants' data and predict the label of the projected data from various participants. The network has the same architecture as the EEGNet model \cite{ref7} and consists of temporal and spatial convolutional neural networks. 
The initial layer is a 2D Convolutional layer with a 1D kernel applied along the temporal dimension, extracting temporal features. Following this, a Depthwise Convolutional layer captures spatial relationships among the channels. Afterward, a separable convolutional layer processes the spatially filtered data, capturing more complex temporal dependencies. Finally, the output is flattened and passed through a dense layer with a softmax activation to produce class probabilities:

\begin{equation}
Y_{\text{features}} = \text{CNN}(Y_{\text{projection}}; W_{\text{Tempconv}}, W_{\text{DWconv}}, W_{\text{Sepconv}})
\tag{3}
\end{equation}

\begin{equation}
Y_{\text{output}} = \text{Softmax}\left(\text{Dense}\left(\text{Flatten}\left(Y_{\text{features}}\right); W'_{\text{dense}}\right)\right)
\tag{4}
\end{equation}

\subsection{Model Training}
\label{model train}
Stochastic batch training was employed for optimizing the model parameters during training.
In each iteration, a random set of samples from multiple participants was selected. During the feed-forward step, each sample was fed into the model, activating the corresponding participant switch.  In the backpropagation step, the mean loss function was calculated for all the samples in the batch. The discriminative network and the dense layer of the selected participants were updated, while the dense layers of other participants remained frozen, as they did not have input data. Algorithm \ref{alg:training} details the complete training procedure for the RISE-iEEG model. 

The model was trained using a cross-entropy loss function and optimized with the Adam algorithm. Early stopping was employed during training to prevent overfitting by monitoring validation accuracy. To further mitigate overfitting, particularly in scenarios with limited sample sizes, L2 regularization was applied to the dense layers of the projection network. The total loss function is defined as:

\begin{equation} 
\mathcal{L}_{\text{total}} = - \frac{1}{M} \sum_{i=1}^{M} \sum_{c=1}^{\text{nb\_classes}} Y_{\text{true}}^{(i, c)} \cdot \log\left(Y_{\text{output}}^{(i, c)}\right) + \lambda \sum_{j=1}^{N} \|W_{\text{Proj}}^{(j)}\|_2^2 \tag{5}
\end{equation}

Model training used TensorFlow 2.2 on a Windows system with dual GTX 1080 GPUs and 32GB RAM, ensuring efficient data processing and calculations.

\begin{algorithm}
\caption{RISE-iEEG Model Training}
\label{alg:training}
\begin{algorithmic}[1]
\REQUIRE Data $\{x_i\}_{i=1}^{N}$, Labels $\{y_i\}_{i=1}^{N}$, Participant Indicator: $S_{\text{i}}$, Projection Networks $\{P_1, \dots, P_N\}$, Discriminative Network $D$, Batch size $B$, Learning rate: $\eta$

\ENSURE Trained model $\{P_1, \dots, P_N, D\}$

\STATE Initialize weights of $P_i$ and $D$
\STATE Set early stopping based on validation accuracy
\FOR{each epoch}
    \FOR{each batch $b$}
        \STATE Sample $B$ examples $(x_i, y_i, S_i)$
        \STATE Initialize batch loss $L^b = 0$
        \FOR{each $(x_i, y_i, S_i)$ in $b$}
            \STATE $z_i = P_{\text{S}_i}(x_i)$, $y_{\text{pred}_i} = D(z_i)$
            \STATE $L^b += L(y_{\text{pred}_i}, y_i)$
        \ENDFOR
        \STATE $L^b = L^b / B + \lambda \sum_i \|W_{P_i}\|_2^2$
        \STATE $W_D = W_D - \eta \nabla L^b$
        \FOR{each participant $i$ in $b$} 
            \STATE $W_{P_i} = W_{P_i} - \eta \nabla L^b$
        \ENDFOR
    \ENDFOR
    \IF{validation accuracy converges}
        \STATE \textbf{break}
    \ENDIF
\ENDFOR
\end{algorithmic}
\end{algorithm}

\subsection{Cross-Validation Paradigms}
\label{data split}
Given that we were working with data from multiple participants and had the participant-specific layers in RISE-iEEG, we established two cross-validation paradigms to assess the model's performance: `same participant' and `unseen participant'. 

In the `same participant' setting, both the training and test sets include data from the same participants, meaning that each participant's data was divided into separate portions for training and testing. We employed pseudo-random selections (folds) to evaluate the model's performance. Within each fold, the data from each participant were divided into training, validation, and test sets. While training the model with the training set, we used the validation set to guide the process and select the best-performing model. Finally, we evaluated the performance of the selected model on the test set.

In the `unseen participant' setting, data from one participant was excluded from the training set to evaluate the model's ability to generalize to new participants. As explained in section \ref{model_archi}, the projection network includes participant-specific dense layers with distinct weights and architectures, due to variations in the positions and numbers of electrodes. Consequently, the projection network needed to be trained using a portion of the data from new participants. To evaluate the model's performance, we employed leave-one-out cross-validation (LOOCV). In each fold, one participant's data served as the test set, while the data from the other participants was used for training and validation. The model was trained in two steps: first, the entire network was trained using data from all participants except one. Next, the projection network was trained using a subset of the new participant's data, while keeping the layers of the discriminative network frozen. During both steps, the validation set was used to guide the process and ensure the selection of the best model based on validation performance. Finally, the model's performance was evaluated using the remaining data from the new participant.

\section{Results}
\subsection{Dataset}
\label{dataset}
In this research, we evaluated RISE-iEEG performance using two publicly available datasets: the Music Reconstruction dataset \cite{ref25} and the AJILE12 dataset \cite{ref64}, both including MNI electrode coordinates for comparison with models like HTNet and enabling neural mechanism interpretation.

\begin{itemize}
    \item \textbf{Music Reconstruction dataset:} This dataset \cite{ref25} comprises electrocorticography (ECoG) recordings from 29 participants listening to rock music, with electrodes placed in the right (11) or left (18) hemisphere based on clinical considerations. The music consisted of 32 seconds of vocals and 2 minutes 26 seconds of instrumentals. For the Singing vs. Music classification task, we divided data into 2-second trials, yielding 16 Singing and 73 Music trials per participant, with the time window duration empirically optimized.
    
    \item \textbf{AJILE12 dataset:} This dataset \cite{ref64} includes ECoG recordings from 12 participants during epilepsy monitoring, with video tracking upper-limb movements. Electrodes were placed in one hemisphere (5 right, 7 left) and varied by participant. The Move vs. Rest classification task used 2-second time windows centered on each trial, yielding at least 150 trials per class for each participant.
\end{itemize}

\subsection{Competing Models}
\label{Comparison models}
We compared the performance of RISE-iEEG with other multi-participant decoders such as HTNet, EEGNet, Random Forest, and Minimum Distance, as implemented in \cite{ref6}. Unlike RISE-iEEG, these decoders require the MNI coordinates of electrodes for each participant.
\begin{enumerate}
    \item \textbf{EEGNet:} EEGNet \cite{ref7} is a convolutional neural network with 2-D convolutional layers to extract spatial and temporal patterns from neural data. In \cite{ref6}, a projection block is added after temporal convolution to map electrode data onto ROIs for decoding across participants. This block uses participant-specific projection matrices, generated by computing radial basis function (RBF) kernel distances between electrodes and brain regions. Therefore, EEGNet requires the MNI coordinates of electrodes to create these matrices.
    \item \textbf{HTNet:} Similar to EEGNet, but with a Hilbert transform layer added after the temporal convolution layer to extract spectral power features \cite{ref6}.
    \item \textbf{Random Forest:} Neural data are projected onto ROIs using the modified EEGNet projection block, enabling the application of the Random Forest classifier across participants' data \cite{ref9}.
    \item \textbf{Minimum Distance:} Like Random Forest, data are projected onto ROIs, and classification is performed using Riemannian mean and distance values \cite{ref51}.
\end{enumerate}


\begin{table}[b]
\centering
\caption{RISE-iEEG performance in the `same participant' setting}
\renewcommand{\arraystretch}{1.3} 
\begin{tabular}{l @{\hspace{1em}} c @{\hspace{4em}} c}
\hline
\  & \multicolumn{2}{c}{\textbf{Tasks}} \\ 
\textbf{Metric} & \textbf{Singing vs. Music} & \textbf{Move vs. Rest}  \\
\hline
\textbf{Precision} & 0.83 {$\pm$} 0.04 & 0.69 {$\pm$} 0.01\\
\textbf{Recall} & 0.85 {$\pm$} 0.02 & 0.69 {$\pm$} 0.01\\
\textbf{F1 score} & 0.83 {$\pm$} 0.03 & 0.68 {$\pm$} 0.01\\
\textbf{AUC score} & 0.88 {$\pm$} 0.07 & 0.76 {$\pm$} 0.03\\
\hline
\end{tabular}
\label{performance same participant}
\end{table}

\subsection{RISE-iEEG Performance in `same participant' Setting}
\begin{figure}[t]
	\centering
	   \includegraphics[width=0.5\textwidth]{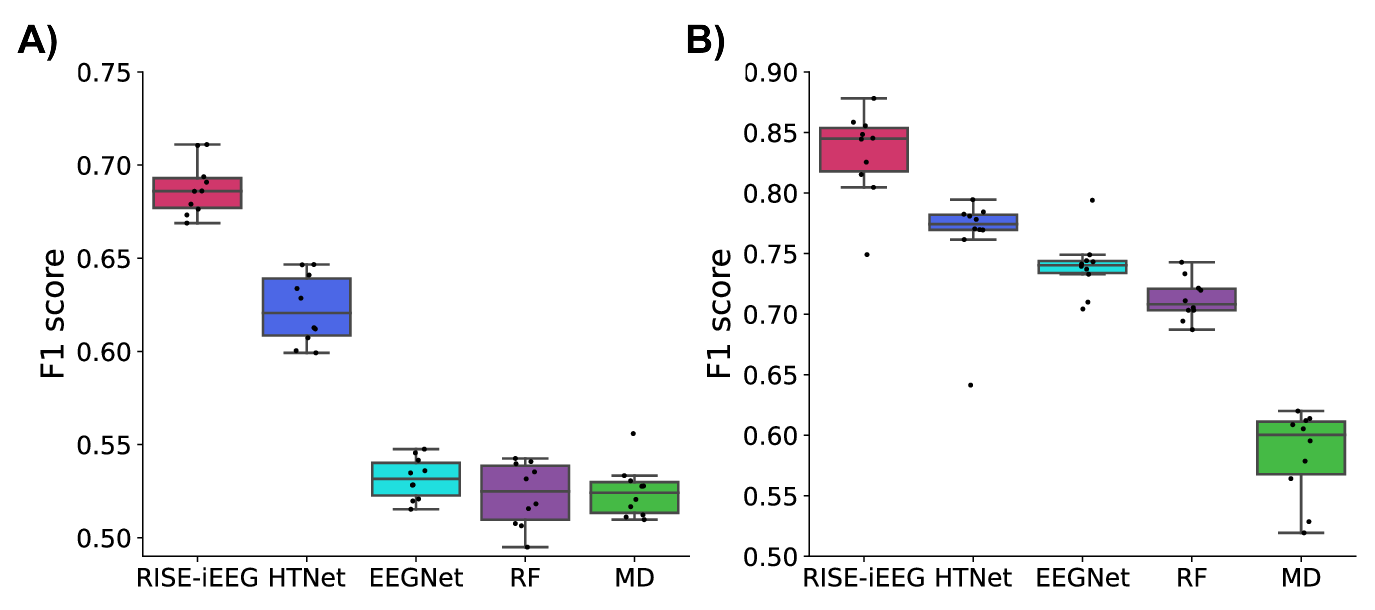}
	\caption{Performance comparison of the RISE-iEEG model (pink) with HTNet (blue), EEGNet (light blue), Random Forest (RF, purple), and Minimum Distance (MD, green) decoders, presented separately for each model in the `same participant' setting for two tasks: (A) the Move vs. Rest task and (B) the Singing vs. Music task. Each point represents the mean F1 score of each fold across participants.}
	\centering
	\label{comparison models in same participant setting}
\end{figure}

In this study, we utilized 10 pseudo-random selections (folds) to assess the performance of the RISE-iEEG model. Within each fold, the training, validation, and test sets comprised 64\%, 16\%, and 20\% of the data from each participant, respectively. The model was trained as described in section \ref{model train}. We evaluated the model's performance on each participant's test set using the F1 score, precision, recall, and AUC score. The performance of the RISE-iEEG model was assessed using two classification tasks, and the results are presented in Tab. \ref{performance same participant}. As shown in the table, the RISE-iEEG model demonstrates acceptable performance in classifying both tasks. We compared the performance of the RISE-iEEG model with advanced decoders, as explained in section \ref{Comparison models}. The results of comparing the performance of these models are illustrated in Fig.\ref{comparison models in same participant setting}. 

Fig.\ref{comparison models in same participant setting} shows that the RISE-iEEG model outperformed other advanced models in both tasks. For the Singing vs. Music classification, RISE-iEEG achieves a test F1 score of 0.83 {$\pm$} 0.03, outperforming HTNet (0.76 {$\pm$} 0.04), EEGNet (0.74 {$\pm$} 0.02), Random Forest (0.71 {$\pm$} 0.02), and Minimum Distance (0.58 {$\pm$} 0.03). In the Move vs. Rest classification, RISE-iEEG attains a test F1 score of 0.69 {$\pm$} 0.01, exceeding HTNet (0.62 {$\pm$} 0.02), EEGNet (0.53 {$\pm$} 0.01), Random Forest (0.52 {$\pm$} 0.02), and Minimum Distance (0.52 {$\pm$} 0.01).

\subsection{RISE-iEEG performance in `unseen participant' setting}
We assessed the model's performance using leave-one-out cross-validation (LOOCV). In each fold, the data from one participant served as the test set, while the network was trained on data from the remaining \(N-1\) participants. The training process involved two steps, as detailed in section \ref{model train}.

\begin{figure}[t]
	\centering
	\includegraphics[width=0.5\textwidth]{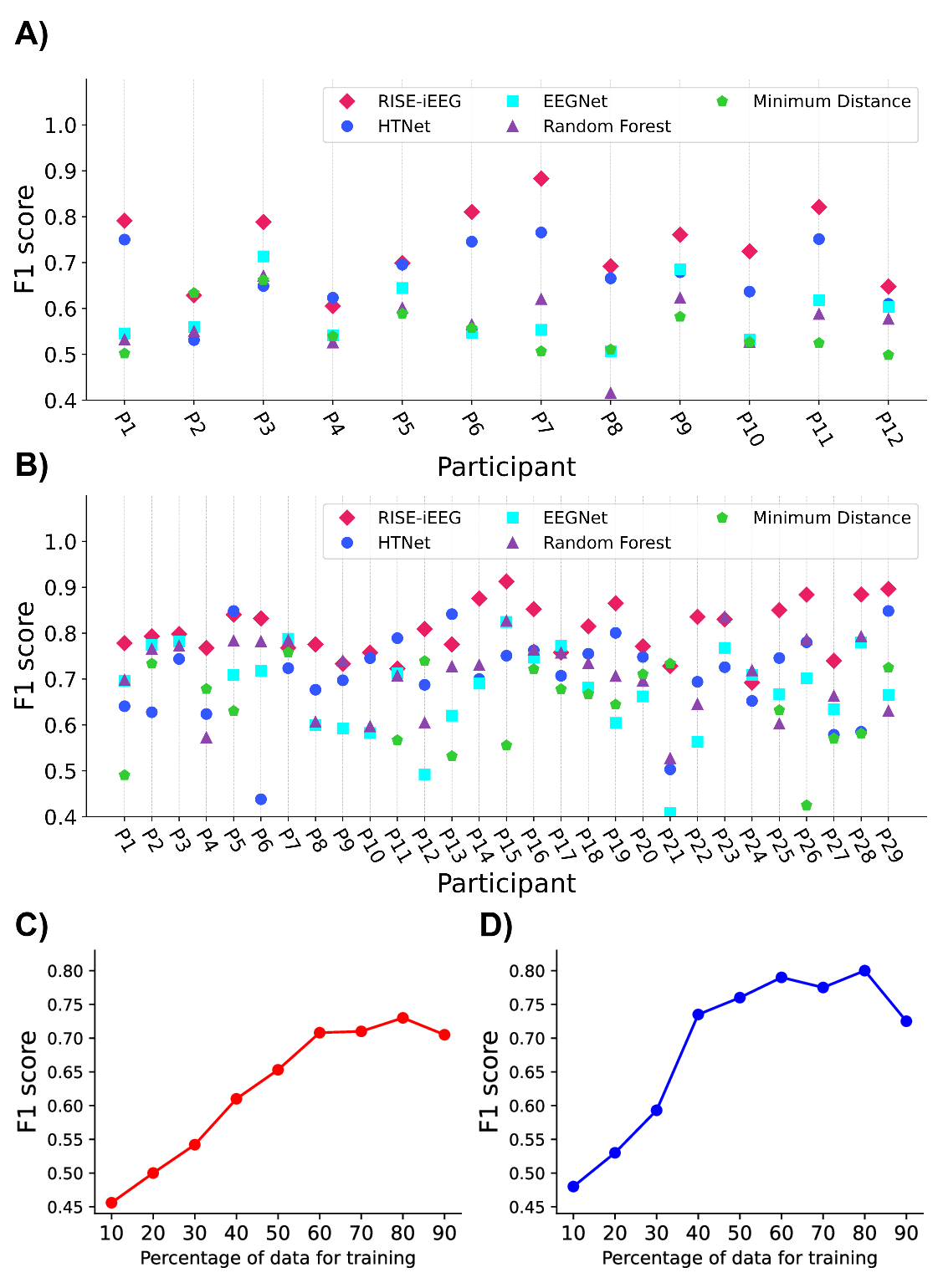}
	\caption{(A, B) Performance comparison of RISE-iEEG (pink) with HTNet (blue), EEGNet (light blue), Random Forest (purple), and Minimum Distance (green) decoders for each participant in the 'unseen participant' setting for the (A) Move vs. Rest and (B) Singing vs. Music tasks. (C, D) Impact of training data amount on model performance in the second training step for the (C) Move vs. Rest  and (D) Singing vs. Music tasks.}
	\centering
	\label{comparison models unseen}
\end{figure}

We evaluated the performance of the RISE-iEEG model against advanced decoders in the `unseen participant' setting. For the Singing vs. Music classification task, RISE-iEEG achieved a test F1 score of 0.80 {$\pm$} 0.05, outperforming HTNet (0.70 {$\pm$} 0.09), EEGNet (0.67 {$\pm$} 0.11), Random Forest (0.71 {$\pm$} 0.08), and Minimum Distance (0.49 {$\pm$} 0.23). In the Move vs. Rest classification, RISE-iEEG attained a test F1 score of 0.74 {$\pm$} 0.08, surpassing HTNet (0.68 {$\pm$} 0.07), EEGNet (0.59 {$\pm$} 0.06), Random Forest (0.57 {$\pm$} 0.06), and Minimum Distance (0.55 {$\pm$} 0.05). The results of these comparisons for each participant individually are illustrated in Fig. \ref{comparison models unseen} (A, B). As shown in this figure, RISE-iEEG notably outperformed other models in classifying data from 20 out of 29 participants in the Singing vs. Music task and from 9 out of 12 participants in the Move vs. Rest task.

To determine the optimal data ratio for training the projection network in the second step, we explored how model performance changes with different ratios of data split between the training and test sets. As shown in Fig.\ref{comparison models unseen} (C, D), Performance improves with larger training data, remains consistent from 60\% to 80\%, but declines at 90\% due to overfitting.

\subsection{Interpretation of Trained RISE-iEEG Model}
\label{IG method}
We used the Integrated Gradient (IG) method \cite{ref11} to identify how stimuli are encoded in the neural activity of different nodes during a task. IG calculates the gradient of the model’s prediction with respect to its input, highlighting the influence of spatiotemporal features on the model’s output.

\begin{figure*}[t]
	\centering
	\includegraphics[width=\textwidth]{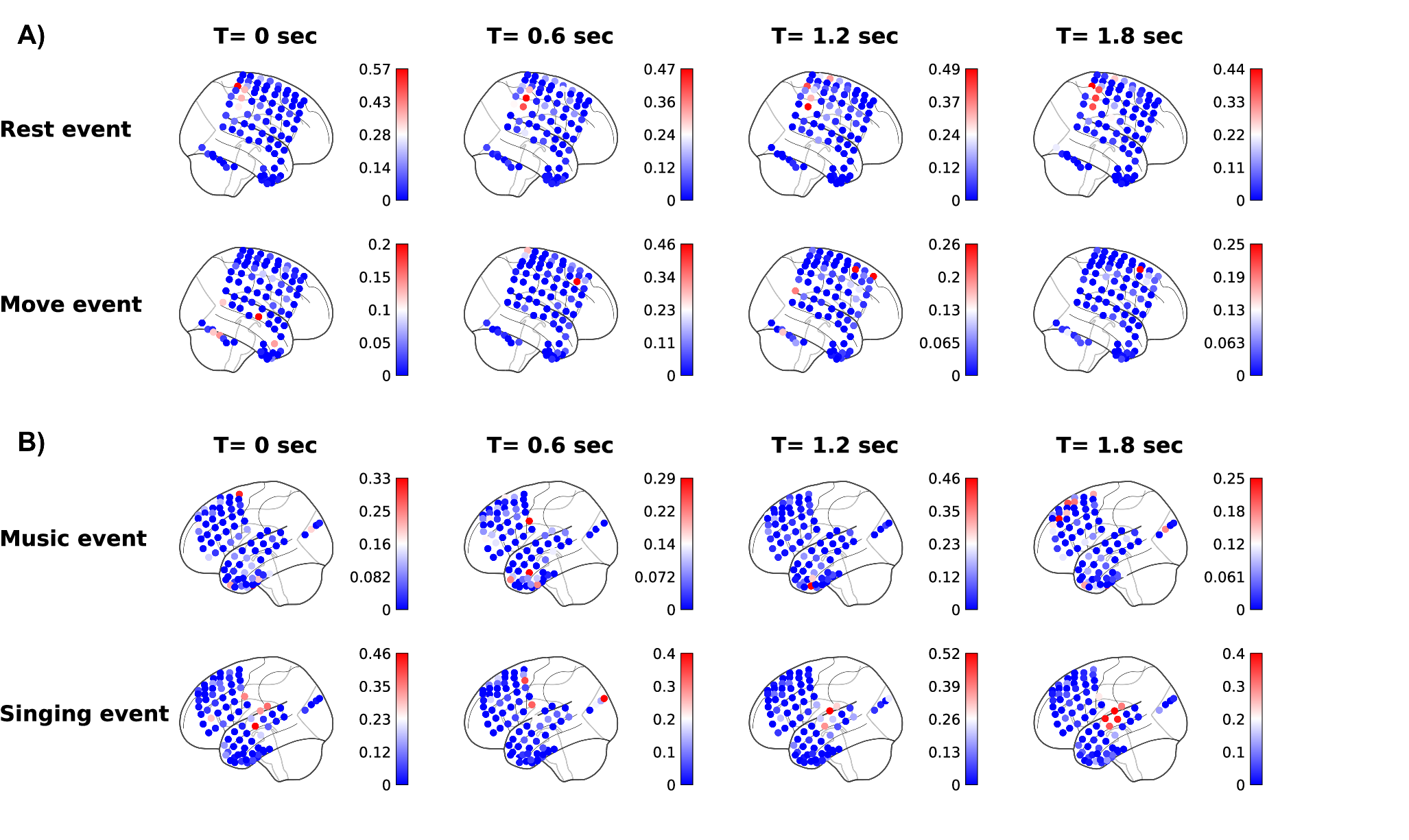}
	\caption{Integrated Gradients (IG) weights for a single participant, showing the significance of each electrode at 600 ms intervals.(A) Move vs. Rest task and (B) Singing vs. Music task. The color scale indicates IG magnitude, with red representing higher importance and blue representing lower importance. This visualization reveals dynamic changes in electrode importance over time, offering insights into engaged brain regions during each task.}
	\centering
	\label{fig. temporal variation IG}
\end{figure*}

To investigate the encoding mechanisms, we performed multiple analyses using the IG method. First, we calculated IG values for all trials of each participant to assess the importance of data from each electrode over time. We analyzed the spatiotemporal variations in mean IG values across participants and presented the results for a single participant, separated by task, in Figure \ref{fig. temporal variation IG}. In this figure, the electrode with the highest IG value signifies its greater contribution to label prediction, indicating that it carries more task-relevant information.

\begin{figure}[t]
	\centering
	\includegraphics[width = 0.5\textwidth]{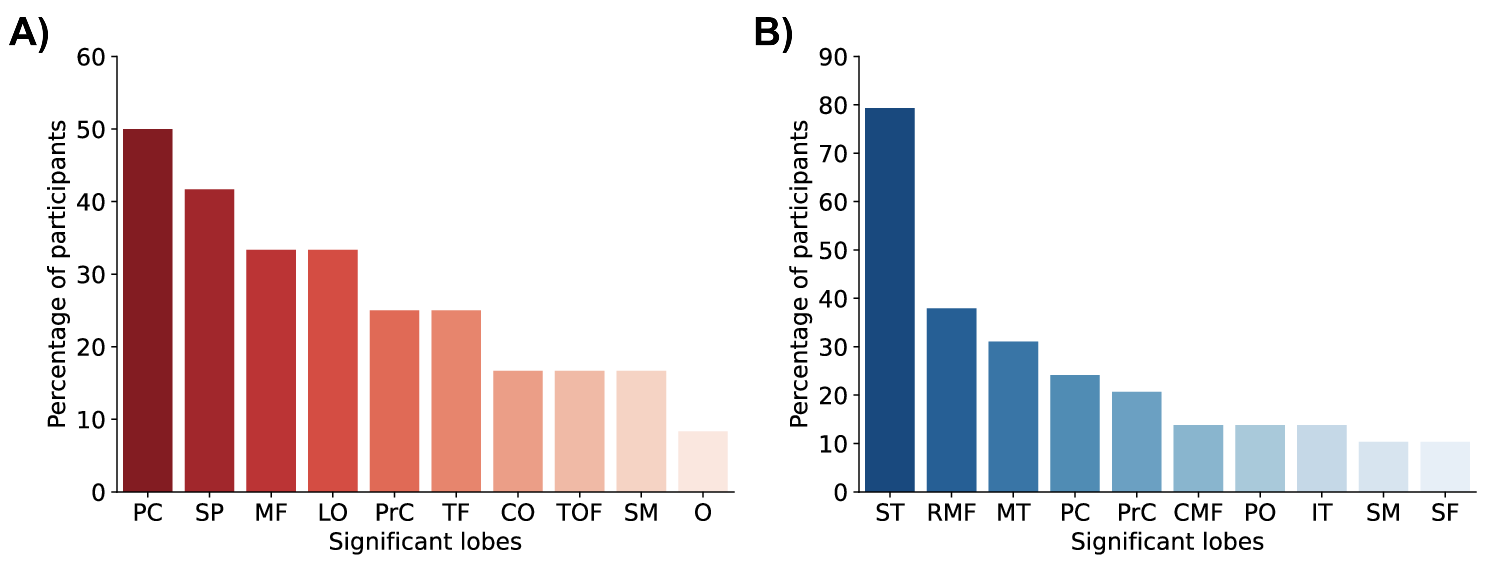}
	\caption{Distribution of significant brain lobes across participants. (A) For the Move vs. Rest task, the significant lobes are as follows: Superior Temporal (ST), Rostral Middle Frontal (RMF), Middle Temporal (MT), Postcentral (PC), Precentral (PrC), Caudal Middle Frontal (CMF), Pars Opercularis (PO), Inferior Temporal (IT), SupraMarginal (SM), and Superior Frontal (SF). (B) For the Singing vs. Music task: Postcentral (PC), Superior Parietal (SP), Middle Frontal (MF), Lateral Occipital (LO), Precentral (PrC), Temporal Fusiform (TF), Central Opercular (CO), Temporal Occipital Fusiform (TOF), SupraMarginal (SM), and Occipital (O).}
	\centering
	\label{fig. histogram significant electrode}
\end{figure}

Fig. \ref{fig. temporal variation IG}(A) shows the variation of IG values for the Move vs. Rest task during the 1-second interval before and after movement onset. As depicted, electrodes in the frontal lobe consistently exhibit high IG weights from 400 ms before to 1 second after movement onset. This observation aligns with the well-established role of the frontal lobe in planning and executing voluntary movements \cite{ref63}. Fig. \ref{fig. temporal variation IG}(B) illustrates the variation of IG values for the Singing vs. Music task during a 2-second trial involving vocal or instrumental music. As shown, the temporal lobe maintains relatively high IG weights across all time intervals, consistent with the primary role of this lobe in auditory processing \cite{ref52}.

Moreover, we identified the significant lobe for each participant ($\text{SL}_{i}^p$). Fig. \ref{fig. histogram significant electrode} illustrates the percentage of participants for whom each lobe is ranked among the top three significant lobes. Only the top three lobes are included in the histogram, as our findings show that, for most participants, these lobes contain at least 80\% of the IG information. Fig. \ref{fig. histogram significant electrode}(A) shows that the Postcentral gyrus is one of the three most significant lobes in 50\% of participants. The proximity of this region to the motor cortex, a key area involved in the planning and execution of voluntary movements, suggests a significant role in motor processing. This finding aligns with previous studies \cite{ref6} conducted on this dataset. As shown in Fig. \ref{fig. histogram significant electrode}(B), the Superior Temporal (ST) lobe ranks among the three most significant lobes in 80\% of participants, indicating that it provides highly informative data for classification in this task. This finding is consistent with prior ECoG studies \cite{ref25}, which highlight the pivotal role of the Superior Temporal lobe in music perception and interpretation.

\begin{figure}[t]
	\centering
	\includegraphics[width = 0.5\textwidth]{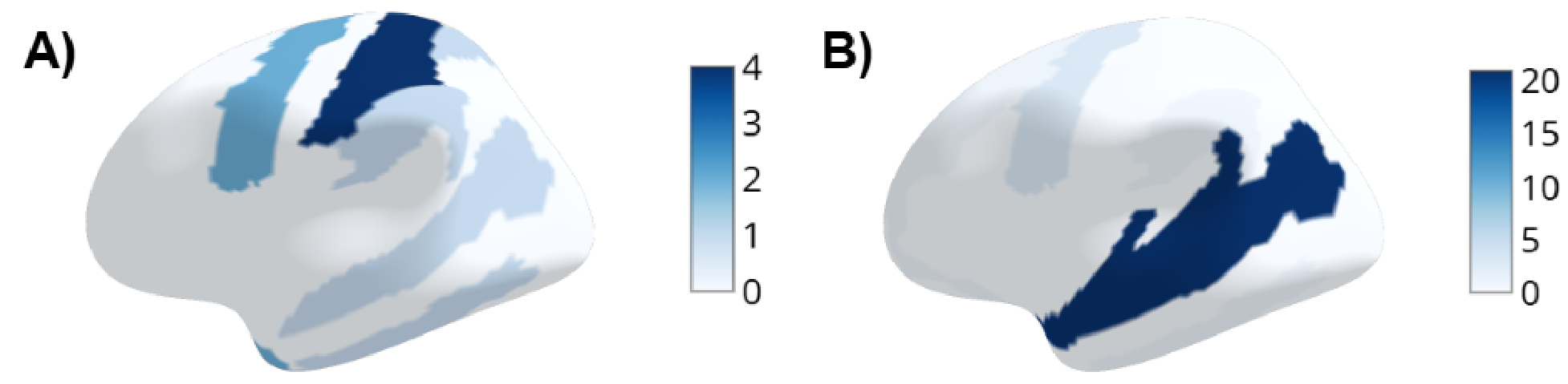}
	\caption{Significant brain regions identified through the analysis of projection network weights, with color intensity indicating each region's importance (A) Move vs. Rest task and (B) Singing vs. Music task.}
	\centering
	\label{hist_table_lobe}
\end{figure}

The Integrated Gradients (IG) method evaluates the contribution of data from each brain region across all layers of the network. In contrast, a more focused analysis can be performed by examining the weights of the participant-specific dense layers within the projection network. This approach enables us to determine the contribution of each brain region in the common space. To identify the most influential brain regions, we analyzed the dense layer weights across all participants and visualized the regions with the highest weights on the brain surface in Fig. \ref{hist_table_lobe}, separated by task. The figure reveals that, for most participants, the Superior Temporal lobe exhibits the highest weights in the Singing vs. Music task, while the Postcentral lobe shows the highest weights in the Move vs. Rest task. These findings align closely with the results obtained using the IG method, further highlighting that the model is capable of identifying task-specific informative brain regions. This interpretability is crucial for understanding the neural mechanisms underlying each task, as it provides insights into how specific brain regions contribute to classification performance.

\section{Discussion}
In this work, we present the RISE-iEEG model, which addresses electrode implantation variability in inter-subject iEEG studies. It consists of a projection network that maps electrode data to a common space using adaptive weights and a discriminative network based on a convolutional neural network (similar to EEGNet) to extract temporal and spatial features from neural data. RISE-iEEG outperformed advanced decoders like HTNet, EEGNet, Random Forest, and Minimum Distance in both `same participant' and `unseen participant' settings across two datasets. This consistent performance demonstrates the model's robustness in handling electrode implantation variability. Additionally, the IG method's results aligned with known physiological principles, validating the model's interpretability.

The projection network in RISE-iEEG has trainable weights, optimizing the mapping process for better feature extraction and improved classification accuracy. In contrast, advanced decoders use fixed projection weights based on electrode-to-ROI physical distance, which misrepresents the brain's functional connectivity and results in lower accuracy.

RISE-iEEG streamlines multi-participant decoding by eliminating the pre-computation required in HTNet, which consists of generating projection matrices and optimizing related hyperparameters. This simplifies decoding and reduces preparation time.

RISE-iEEG achieves higher prediction accuracy but has two key limitations. First, it requires fine-tuning the projection network with a portion of data from new participants. Second, it has more parameters than HTNet due to the trainable weights of the projection network, posing challenges for small datasets. However, applying L2 regularization enables RISE-iEEG to perform comparably to HTNet and EEGNet with similar data sizes, eliminating the need for larger datasets. Future research could explore unsupervised learning for the projection network to avoid fine-tuning with new data and incorporate time-varying weights in the projection network to enhance dynamic adaptation to neural activity and complex interactions

\section{Conclusion}
In this study, we introduced the RISE-iEEG model, a generalizable neural decoder for iEEG data from diverse experiments. It includes a participant-specific projection network that maps data into a shared low-dimensional space, followed by a discriminative deep neural network. This approach addresses inter-subject electrode variability without requiring MNI coordinates. RISE-iEEG outperforms advanced models like HTNet and EEGNet across two datasets, showcasing its effectiveness. Additionally, it reveals neural encoding mechanisms, identifying the Superior Temporal and Postcentral lobes as key nodes in the Music Reconstruction and AJILE12 datasets. The RISE-iEEG model demonstrates robust decoding performance and offers an interpretable architecture that supports the exploration of neural mechanisms, making it a valuable tool for advancing neuroscience research.

\section*{Code and data availability}
\begin{sloppypar}
The RISE-iEEG code is available at \url{https://github.com/MaryamOstadsharif/RISE-iEEG.git} and works with the publicly accessible Music Reconstruction dataset (\url{https://zenodo.org/records/7876019}) and AJILE12 dataset (\url{https://dandiarchive.org/dandiset/000055/0.220127.0436}), allowing full reproduction of the study's findings and figures.
\end{sloppypar}

\appendices

\bibliographystyle{IEEEtran}
\bibliography{main_ref}

\end{document}